\documentclass[conference, a4paper]{APSIPA2021}
\usepackage{amsmath}
\usepackage{multirow}
\usepackage{threeparttable}

\usepackage{amssymb}
\usepackage{subcaption}
\usepackage{graphicx}
\usepackage{cite}
\usepackage{url}

\usepackage{geometry}
\usepackage{fancyhdr}

\fancypagestyle{firststyle}{
  \fancyhf{}
  \fancyhead[C]{2024 Asia Pacific Signal and Information Processing Association Annual Summit and Conference (APSIPA ASC)}
}

\begin{document}

\title{Enhancing Security Using Random Binary Weights\\in Privacy-Preserving Federated Learning}

\author{
\authorblockN{
Hiroto Sawada\authorrefmark{1},
Shoko Imaizumi\authorrefmark{2}, and 
Hitoshi Kiya\authorrefmark{3}
}

\authorblockA{
\authorrefmark{1}
Graduate School of Science and Engineering, Chiba University, Chiba, Japan 
E-mail: 24wm3202@student.gs.chiba-u.jp  
}

\authorblockA{
\authorrefmark{2}
Graduate School of Engineering, Chiba University, Chiba, Japan 
E-mail: imaizumi@chiba-u.jp  
}

\authorblockA{
\authorrefmark{3}
Faculty of System Design, Tokyo Metropolitan University, Tokyo, Japan 
E-mail: kiya@tmu.ac.jp  
}
}

\maketitle
\thispagestyle{firststyle}
\pagestyle{fancy}

\begin{abstract}
In this paper, we propose a novel method for enhancing security in privacy-preserving federated learning using the Vision Transformer. In federated learning, learning is performed by collecting updated information without collecting raw data from each client. However, the problem is that this raw data may be inferred from the updated information.
Conventional data-guessing countermeasures (security enhancement methods) for addressing this issue have a trade-off relationship between privacy protection strength and learning efficiency, and they generally degrade model performance. In this paper, we propose a novel method of federated learning that does not degrade model performance and that is robust against data-guessing attacks on updated information. In the proposed method, each client independently prepares a sequence of binary (0 or 1) random numbers, multiplies it by the updated information, and sends it to a server for model learning. In experiments, the effectiveness of the proposed method is confirmed in terms of model performance and resistance to the APRIL (Attention PRIvacy Leakage) restoration attack.
\end{abstract}

\pagestyle{empty}

\section{Introduction}

The rapid development of AI technology has accelerated the growth of businesses and services that use deep learning. 
However, training a model with many parameters requires a large amount of training data, and preparing this training data takes a lot of time and effort. 
Additionally, we have to take into account the privacy information contained in the training data.
In addressing this issue, federated learning, which is a distributed learning method, has attracted much attention \cite{c1,c1_2,c1_3}.

\begin{figure}[tb]
\begin{center}
\includegraphics[width=8.2cm]{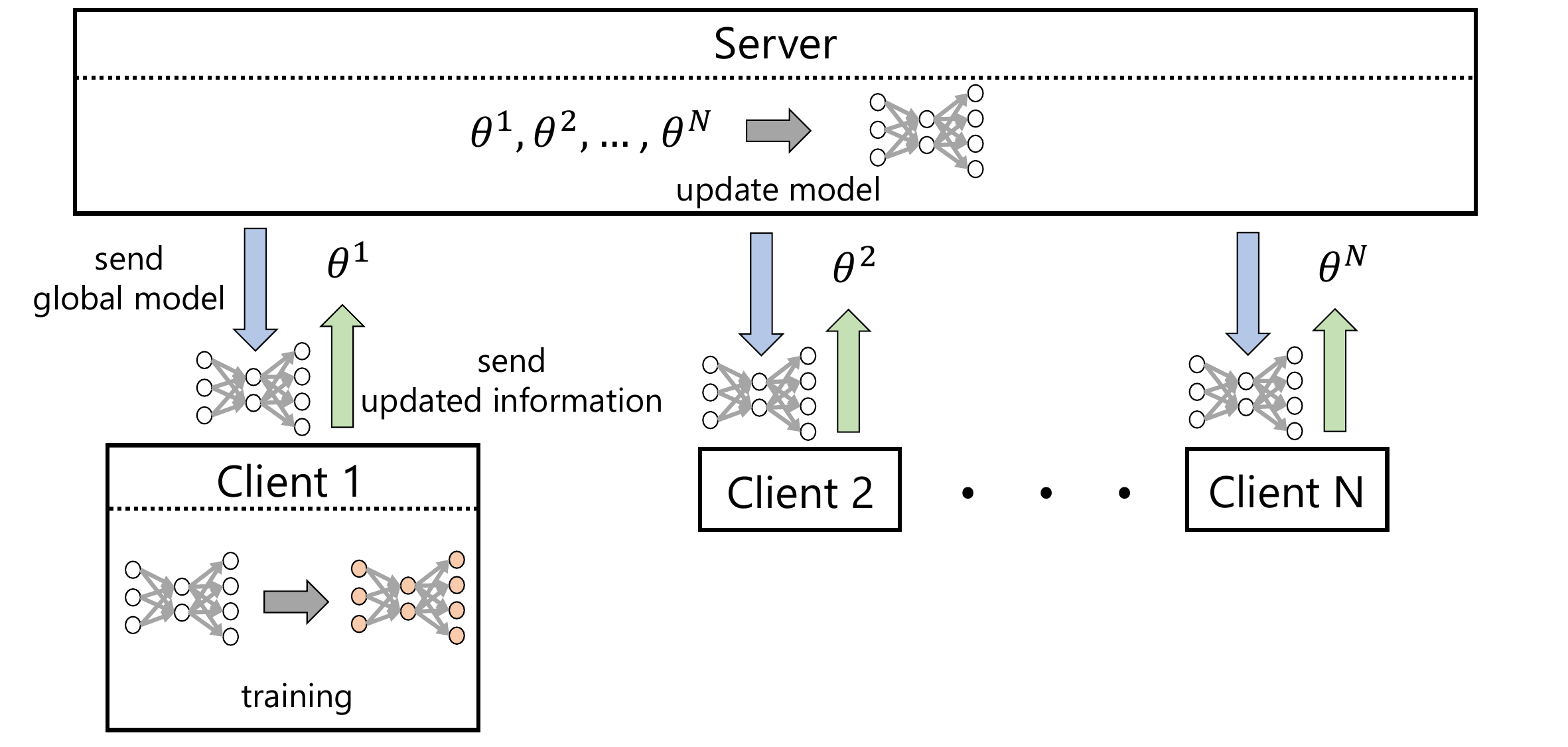}
\end{center}
\caption{Overview of federated learning}
\label{fig:0}
\end{figure}

In federated learning, multiple parties cooperate for deep learning, as shown in Fig. \ref{fig:0}.
Here, we define a server as a party that provides a model and clients as parties that have training data.
Each client independently trains the model shared by the server using its own training data.
The clients send their updated information obtained through the training to the server, and the server integrates all the information to update the model.
As described above, federated learning can efficiently train high-performance models with a large number of parameters.
However, attacks \cite{c2,c3,c4,c5} have been proposed to infer training data on the basis of such updated information.

For this reason, security enhancement methods have been actively researched to protect privacy against such attacks.
One method \cite{c7} can compute in the encryption domain by using homomorphic cryptography \cite{c6} and retain accuracy; however, it is computationally expensive.
%
In another method \cite{c9}, updated information is concealed by combining model subsampling, model shuffling, and blanket noise.
The method is, however, vulnerable to collusion between a server and a party shuffling the model.
Consequently, this approach causes a reduction in accuracy.
There is a different type of method based on differential privacy \cite{c10,c11,c12,c13,c14}. 
In this type, each client adds a random number sequence following a specific distribution to the updated information.
To enhance security, however, the variance in the distribution has to be enlarged. 
Thus, the model performance tends to decrease.

There also exists research on the use of the Vision Transformer (ViT) \cite{c15} for federated learning.
ViT is an image classification model known for its high performance.
However, a gradient leakage attack called Attention Privacy Leakage (APRIL) \cite{c5}, which attempts to recover training images on the basis of updated information for ViT, has become a serious problem.
Against APRIL, Aso et al. proposed an effective model encryption method \cite{c15_2}.
This method maintains accuracy but does not possess resistance against attacks from internal parties such as clients.
Lu et al. locked the positional embedding layer of ViT to prevent it from being updated \cite{c5}.
Although this method is robust against internal attacks, the performance is degraded when using a model without pre-training.

To tackle the above issue, we propose a novel security enhancement method for ViT-based federated learning without performance deterioration. 
In the proposed method, each gradient of ViT is multiplied by random binary weights, and all parameters including those in the positional embedding layer can be updated.
Simulation results show the effectiveness of our method through an evaluation on the classification accuracy and resistance against APRIL.

\section{Preparation}
We enhance security for federated learning with ViT, assuming APRIL as a gradient leakage attack. 
In this section, we first give brief explanations of ViT and APRIL, and we then summarize several previous methods that have resistance against gradient leakage attacks. 

\subsection{Vision Transformer}
ViT with self-attention mechanisms has attracted much attention for its high performance in the field of image recognition and classification \cite{c15}. 
On the other hand, in the federated learning field, there have been studies on attacks targeting ViT, and a powerful restoration attack called APRIL has been proposed \cite{c5}.

\begin{figure}[t]
\begin{center}
\begin{minipage}[b]{0.9\linewidth}
\begin{center}
\includegraphics[width=7.7cm]{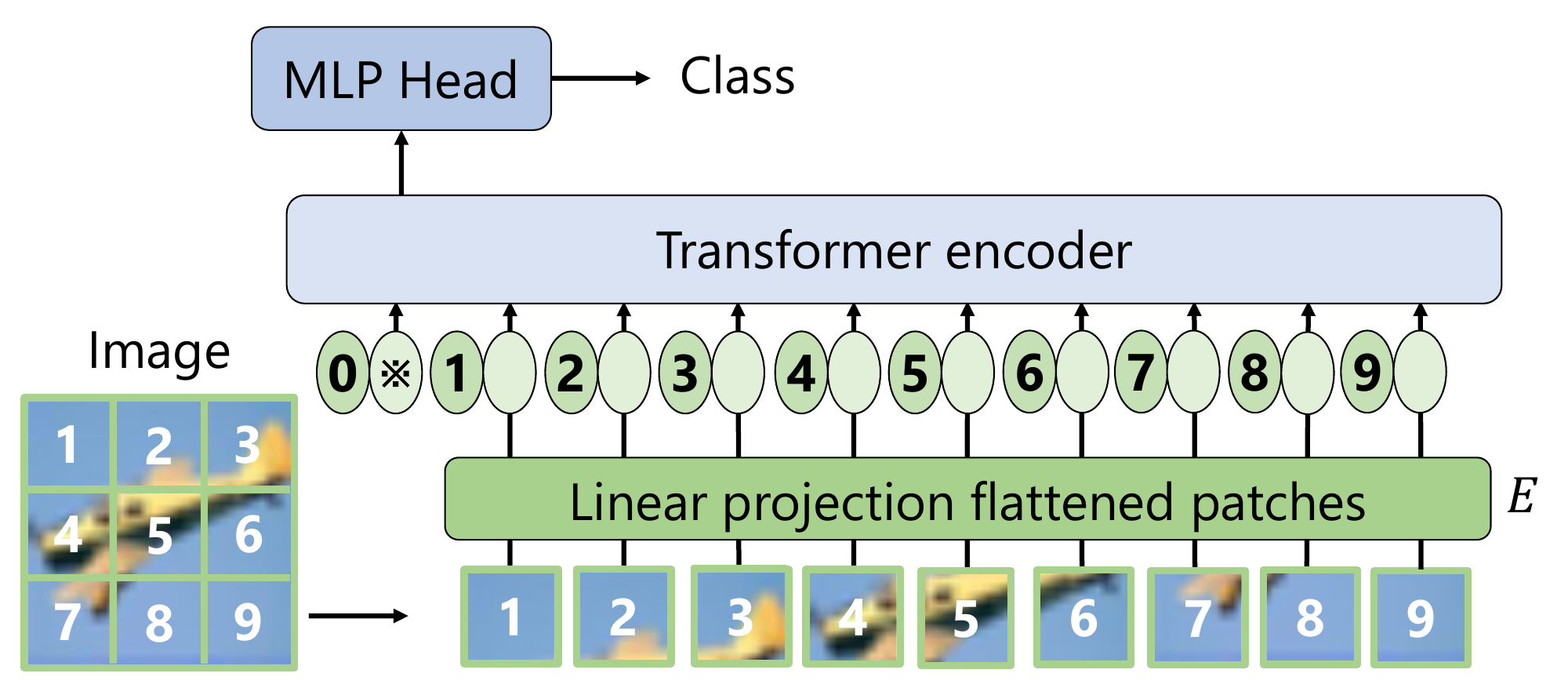}
\end{center}
\subcaption{Overview}
\label{fig:2_1_ViT}
\end{minipage}
\end{center}
\hspace*{45mm}
\begin{center}
\begin{minipage}[b]{0.5\linewidth}
\begin{center}
\includegraphics[width=3cm]{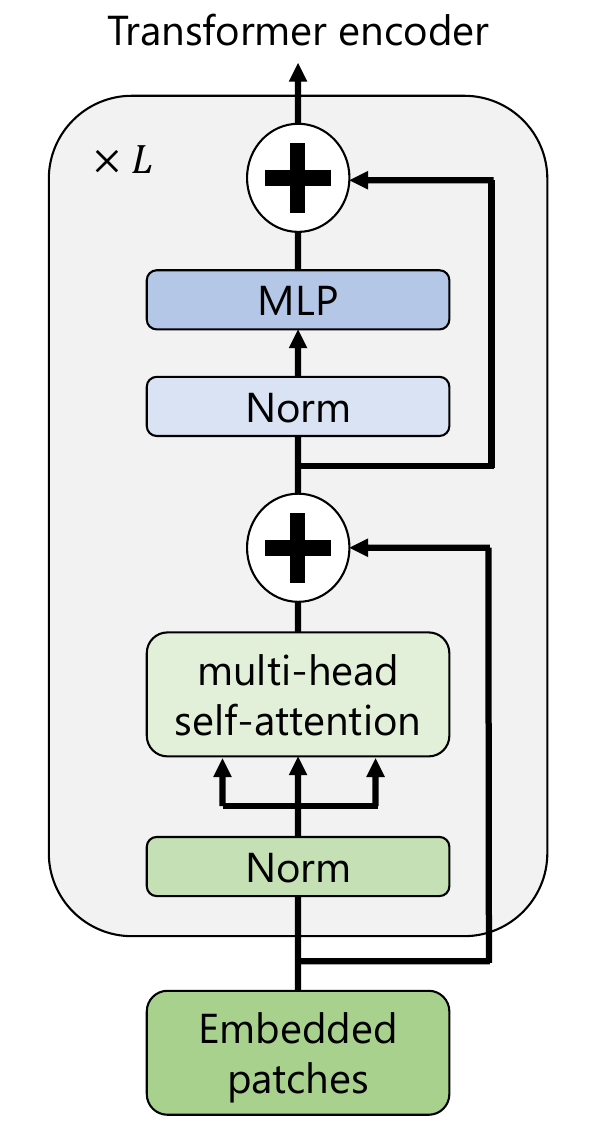}
\end{center}
\subcaption{Transformer encoder layer}
\label{fig:2_2_ViT_TE}
\end{minipage}
\end{center}
\caption{ViT structure}
\label{fig:2_ViT}
\end{figure}

Fig. \ref{fig:2_ViT}\subref{fig:2_1_ViT} illustrates the procedure for ViT.
ViT first divides an input image $x\in\mathbb{R}^{H \times W \times C}$ into patches $x_{p}^s\in\mathbb{R}^{P^{2} \times C}$, where $H$, $W$, and $C$ are the height, width, and number of channels.
$S, s \in \{ 1, 2, \cdots , S\}$ and $P$ are the number of patches, the number assigned to a patch, and the height and width of a patch, respectively. 
A linear layer $E\in\mathbb{R}^{(P^{2} \times C) \times D}$ then dimensionally transforms each patch to be adequate for the Transformer encoder layer.
Note that $D$ represents the vector length after embedding. 
In addition, a class token $x_{class}\in\mathbb{R}^{D}$ is set at the beginning of the patches to represent the features of the entire image. 
Next, the positional embedding layer $E_{pos}\in\mathbb{R}^{(S+1) \times D}$ embeds information on the location relationship among class tokens and patches. 
The calculation result $z_{0}\in\mathbb{R}^{(S+1) \times D}$ is given by
\begin{equation}
z_{0}=[x_{class} ; x^{1}_{P} E ; x^{2}_{P} E ; \cdots ;x^{S}_{P}E] + E_{pos},
\label{eq_z0}
\end{equation}
and it is input to the Transformer encoder layer.

As shown in Fig. \ref{fig:2_ViT}\subref{fig:2_2_ViT_TE}, the Transformer encoder layer is further decomposed into three layers: layer normalization (Norm), multi-head self-attention (MSA), and multi-layer perceptron (MLP).
Let $z_{l-1}$ be the input and $z_{l}$ be the output in the Transformer encoder layer in the $l \in \{ 1, 2, \cdots , L\}$-th layer.
We explain the calculation process for obtaining $z_{l}$ from $z_{l-1}$ in the $l$-th Transformer encoder layer.
First, $z_{l-1}$ is normalized by LN.
In LN, the output $LN(z_{l-1})$ is obtained by
\begin{equation}
LN(z_{l-1})=\gamma \frac{z_{l-1}-E[z_{l-1}]}{\sqrt{Var[z_{l-1}]+\epsilon}} + \beta ,
\label{eq_ln}
\end{equation}
where $\gamma$ and $\beta$ are learnable parameters, and $\epsilon$ is a small constant for ensuring that the denominator never takes zero. 
Next, MSA carries out computation to obtain the output $SA(LN(z_{l-1}))$ by using multiple self-attentions (SAs): 
\begin{equation}
SA(LN(z_{l-1})) = softmax\left(\frac{q_{l}k_{l}^{T}}{\sqrt{D_{h}}}\right) v_{l}.
\label{eq_SA}
\end{equation}
In the computation of SA, $\left( q_l, k_l, v_l \right)$ for the input $LN(z_{l-1})$ are determined by
\begin{equation}
\begin{aligned}
\begin{cases}
q_l = LN(z_{l-1}) U_{ql}, \\
k_l = LN(z_{l-1}) U_{kl}, \\
v_l = LN(z_{l-1}) U_{vl}.
\end{cases}
\end{aligned}
\label{eq_qkv}
\end{equation}
Here, $\left( U_{ql}, U_{kl}, U_{vl} \right) \in \mathbb{R}^{D \times D_h} $ are learnable matrices in the $l$-th SA layer, and $D_{h}$ is the vector length $D$ divided by the number of heads.
Additionally, $\left( q_l, k_l, v_l \right)$ are the input queries, keys, and values linearly transformed by $\left( U_{ql}, U_{kl}, U_{vl} \right)$.
From \eqref{eq_SA}, in MSA, the output $MSA(LN(z_{l-1}))$ is given by
\begin{equation}
MSA(LN(z_{l-1}))= \begin{bmatrix}
SA_{1}; SA_{2}; \cdots ; SA_{T},
\end{bmatrix} U_{msa},
\label{eq_MSA}
\end{equation}
where $SA_{t}$ is the output of the $t \in \{ 1, 2, \cdots , T\}$-th SA, and $U_{msa} \in \mathbb{R}^{k \cdot D_{h} \times D}$ is a matrix for returning the shape of the MSA output to that of the input.
According to \eqref{eq_MSA}, in Transformer encoder layer, the output $z_{l}$ is obtained by
\begin{equation}
\begin{aligned}
z_{l} = &MLP(LN(z_{l}')) + z_{l}',\\
   &where\ \  z_{l}' = MSA(LN(z_{l-1})) + z_{l-1}.
\label{eq_MLP}
\end{aligned}
\end{equation}

Finally, we obtain the output $y$ of ViT, which can be calculated from LN by inputting only the class token $z_{l}^0$ from the outputs of the $L$-th Transformer encoder layer.
\begin{equation}
y = LN(z_{l}^0).
\label{eq_fin_ViT}
\end{equation}
ViT predicts the label of the input image on the basis of the output $y$.

\subsection{Attention Privacy Leakage}

In federated learning, methods for updating a model can be broadly classified into two categories: Federated Stochastic Gradient Descent (FedSGD) and Federated Averaging (FedAvg) \cite{c1}.
As mentioned above, there is a gradient leakage attack called APRIL \cite{c5} that uses updated information to infer input images.
APRIL targets FedSGD-based learning using ViT.
We summarize the attack below.

First, attackers calculate the input to the Transformer encoder layer $z_{0}$:
\begin{equation}
\begin{aligned}
&\frac{\partial l}{\partial z_{0}} z_{0}^{T} = U_{q1}^{T} \cdot \frac{\partial l}{\partial U_{q1}} + U_{v1}^{T} \cdot \frac{\partial l}{\partial U_{v1}} + U_{k1}^{T} \cdot \frac{\partial l}{\partial U_{k1}},\\
&\frac{\partial l}{\partial z_{0}} = \frac{\partial l}{\partial E_{pos}}.
\label{eq_APRIL_z0_2}
\end{aligned}
\end{equation}
Here, $\frac{\partial l}{\partial \ast}$ is the gradient of the parameter $\ast$ on the loss function.
Next, the inferred image $x'$ is obtained by inversion of \eqref{eq_z0}:
\begin{equation}
x' = E \times (z_{0} - E_{pos})^{T}.
\label{eq_APRIL_x}
\end{equation}
In \eqref{eq_APRIL_x}, $E$ and $E_{pos}$ refer to the parameters of the global model before being updated.
APRIL attempts to reconstruct the input images for learning by using the above procedure.

\subsection{Previous Method}

\begin{figure}[t]
\begin{center}
\begin{minipage}[b]{0.9\linewidth}
\begin{center}
\includegraphics[width=8cm]{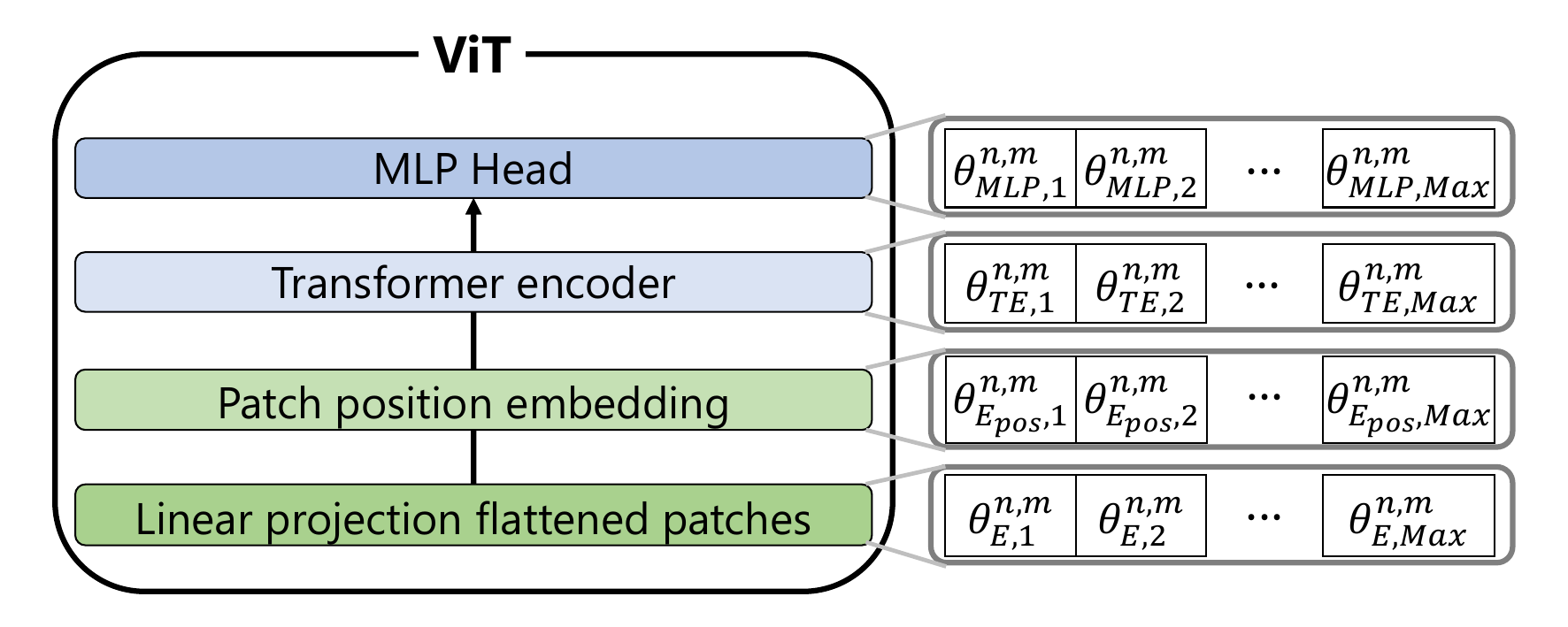}
\end{center}
\subcaption{Gradients with normal learning}
\label{fig:3_1_no_enc}
\end{minipage}
\end{center}
\begin{center}
\begin{minipage}[b]{0.9\linewidth}
\begin{center}
\includegraphics[width=8cm]{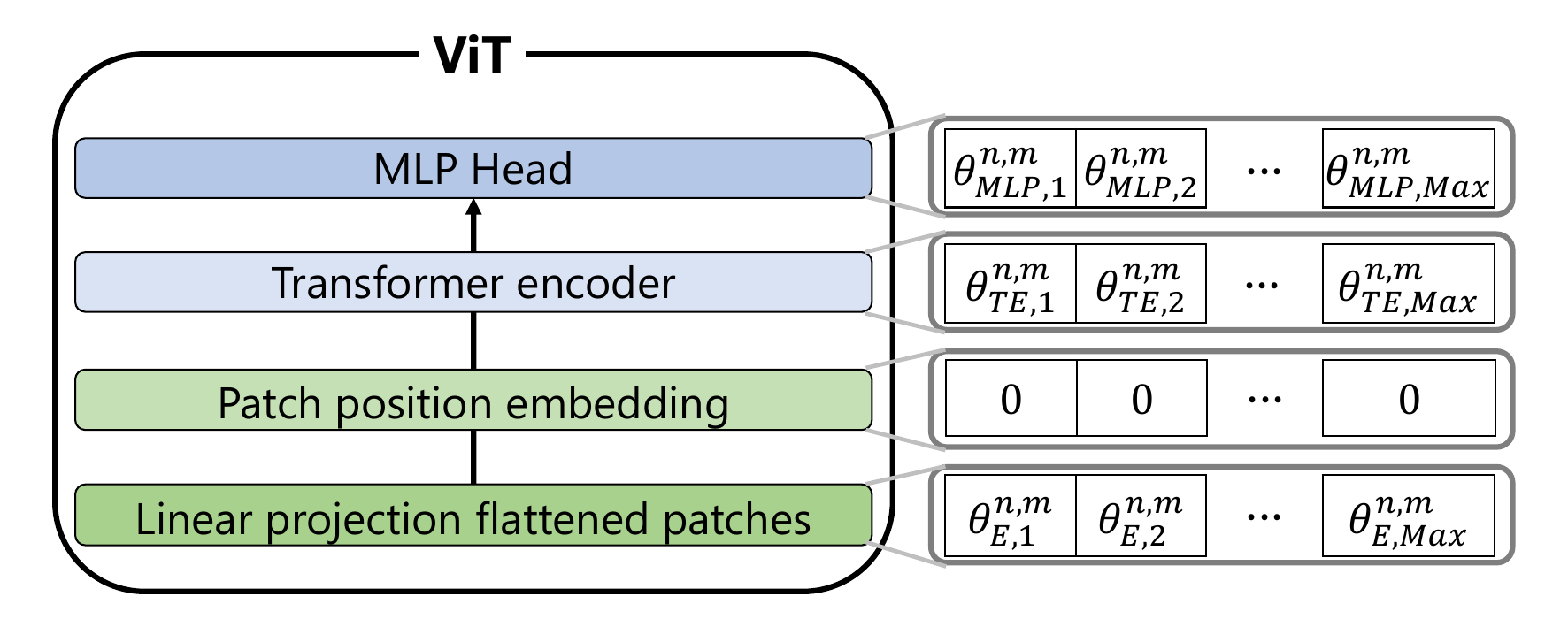}
\end{center}
\subcaption{Gradients with fixed-position method\cite{c5}}
\label{fig:3_2_j_enc}
\end{minipage}
\end{center}
\begin{center}
\begin{minipage}[b]{0.9\linewidth}
\begin{center}
\includegraphics[width=8cm]{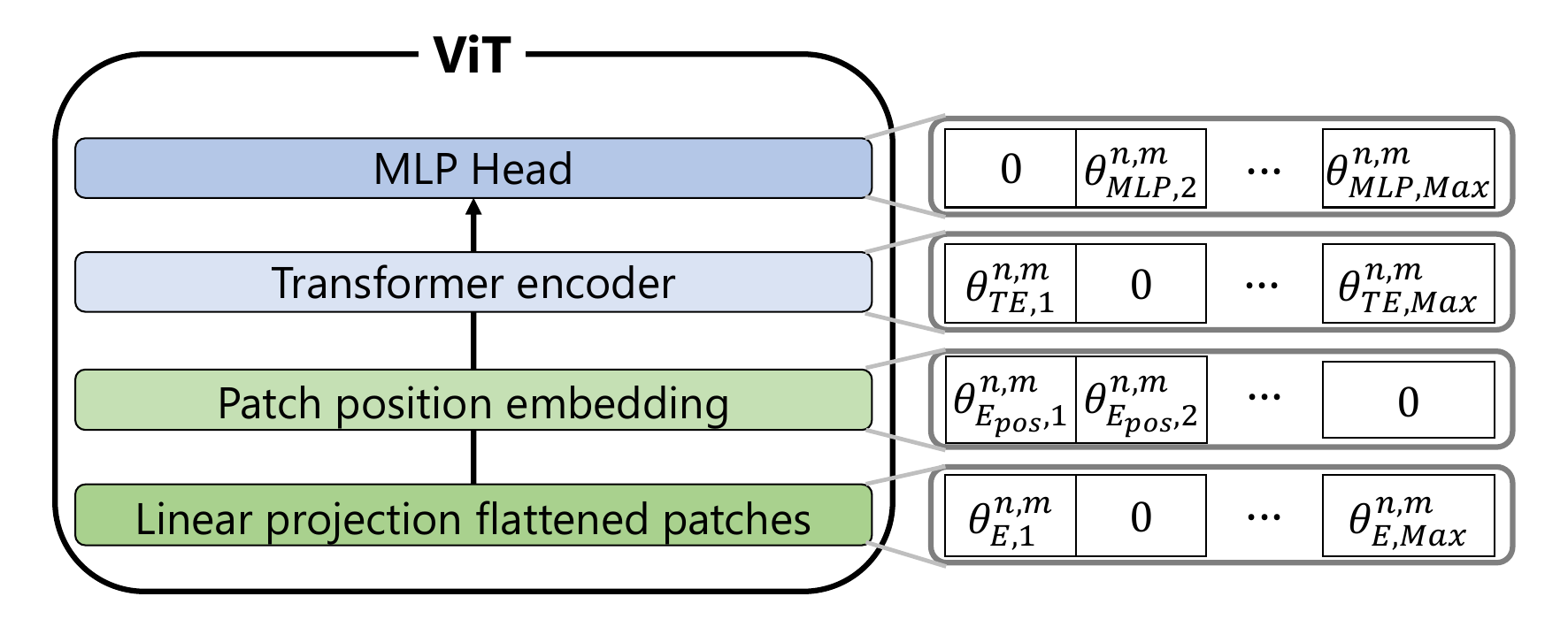}
\end{center}
\subcaption{Gradients with random binary weight (proposed)}
\label{fig:3_3_p_enc}
\end{minipage}
\end{center}
\caption{Update of gradients}
\label{fig:3_grad}
\end{figure}

Several security enhancement methods for federated learning have been proposed.
A representative method is differential privacy \cite{c10,c11,c12,c13,c14}.
In differential privacy, we add noise to the updated information so as to enhance security at a low computational cost.
However, differential privacy degrades model performance.
Another method lets a third party scramble the updated information for high anonymity for each client \cite{c9}.
This approach requires the assumption that the third party is absolutely credible.

In contrast, there is another security enhancement method for federated learning with ViT, which is called the fixed-position method \cite{c5}.
This method is implemented assuming FedSGD.
In FedSGD, the update of parameters is expressed by 
\begin{equation}
\label{eq_FedSGD_1}
w_{A, i}^{m+1} = w_{A,i}^{m} - \eta \frac{\sum_{n=1}^N \theta_{A, i}^{n, m}}{N},
\end{equation}
where $w_{A, i}^m+1$ and $w_{A, i}^m$ are the $i$-th global model parameters of the $A$-th layer after the $m+1$-th or $m$-th updates, $\eta$ is the learning rate, $N$ is the number of clients, and $\theta_{A, i}^{n, m}$ is the gradient for updating $w_{A, i}^m$ calculated by the $n$-th client.
The parameter update is carried out for each batch training.
Fig. \ref{fig:3_grad}\subref{fig:3_1_no_enc} depicts the gradients obtained from training using a plain FedSGD.
In this figure, $Max$ is the number of gradients in each layer.

The fixed-position method enhances security but does not learn the positional embeddin layer.
Fig. \ref{fig:3_grad}\subref{fig:3_2_j_enc} shows gradients transmitted in federated learning with the fixed-position method.
All the gradients in the positional embedding layer are changed to 0.
In this case, the learning process is described as follows.

\begin{quote}
\renewcommand{\labelenumi}{\bf Step\ \theenumi:}
\begin{enumerate}
\setlength{\leftskip}{20pt} 
{\item \noindent
A server sends global model information to each client.

\item \noindent
Each client trains each local model using his/her own dataset.

\item \noindent
Each client converts gradients in the positional embedding layer to 0.

\item \noindent
Each client sends the updated gradients to the server.

\item \noindent
The server integrates the information sent from all the clients and updates the global model.
}
\end{enumerate}
\end{quote}
We repeat the steps a specified number of times. 
The fixed-position method is highly robust against APRIL.
The attacker can be a server, one of the clients, or an external third party; the security enhancement method is effective against all of them.
The method, however, changes all gradients in the positional embedding layer to zero, so we cannot update the parameters.
This leads to severe degradation in model performance when learning a model without pre-training.

\section{Proposed Method}

We propose a novel method that enhances security for federated learning using ViT.
The proposed method is robust against APRIL without affecting model performance.

\subsection{Main Procedure}

The following series of steps is the main procedure for the federated learning to which the proposed method is applied.
\begin{quote}
\renewcommand{\labelenumi}{\bf Step\ \theenumi:}
\begin{enumerate}
\setlength{\leftskip}{20pt}
{\item \noindent
A server sends global model information to each client.

\item \noindent
Each client trains each local model using his/her own dataset.

\item \noindent
Each client multiplies gradients by a random binary sequence.

\item \noindent
Each client sends the updated gradients to the server.

\item \noindent
The server integrates the information sent from all the clients and updates the global model.
}
\end{enumerate}
\end{quote}
We repeat this procedure a predefined number of times.
The main difference from the previous method \cite{c5} is Step 3.
Next, we explain Steps 3 and 5 in more detail.

\subsection{Multiplication of Random Binary Weights}

Fig. \ref{fig:3_grad}\subref{fig:3_3_p_enc} illustrates a simplified gradient processing for each layer of ViT in the proposed method.
In Step 3, each client multiplies gradients by random binary weights $B_{A, i}^{n, m} \in \{ 0, 1 \}$.
Here, a weighted gradient $\theta_{A, i}^{n, m'}$ is given by
\begin{equation}
\theta_{A, i}^{n, m'} = B_{A, i}^{n, m} \times \theta_{A, i} ^{n, m},
\label{eq:randomB}
\end{equation}
where $B_{A, i}^{n, m}$ is a binary weight that is multiplied by $\theta_{A, i}^{n, m}$.
In the case that the occurrence probability of zeros $R \in  \lbrack0, 1 \rbrack$ is set higher, more robust privacy protection would be provided; however, the model may not be updated correctly.
We tackle this issue by changing the random binary weights $B_{A, i}^{n, m}$ at each epoch.

It is possible to define a different probability for each layer. 
The positional embedding layer $E_{pos}$ and linear layer $E$ are layers that include the major information of an image, so APRIL uses these layers for image restoration.
Thus, for instance, it would be useful to assign high probabilities for $E_{pos}$ and $E$ while assigning low probabilities for other gradients.
In this paper, we apply the same probabilities $R$ to all layers to simplify the argument.

\subsection{Gradient Integration}

Next, we elaborate on the gradient integration in Step 5.
The proposed method integrates gradients on the basis of FedSGD.
However, it is not appropriate to directly apply \eqref{eq_FedSGD_1} to the proposed method because each client randomly converts the gradients to 0.
The gradient integration is implemented as:
\begin{equation}
  w_{A, i}^{m+1}=
  \begin{cases}
    w_{A,i}^{m} - \eta \frac{\sum_{n=1}^N \theta_{A, i}^{n, m'}}{\sum_{n=1}^{N} B_{A, i}^{n, m}} & \text{if $0 < \sum_{n=1}^{N} B_{A, i}^{n, m} \leq N$} \\
    w_{A, i}^{m}                 & \text{if $\sum_{n=1}^{N} B_{A, i}^{n, m} = 0$}. \\
  \end{cases}
\label{eq_FedSGD_t}
\end{equation}
Note that zero gradients are excluded when calculating the mean of the gradients.
As an exception, parameters cannot be updated in the case of $\sum_{n=1}^{N} B_{A, i}^{n, m} = 0$.


\section{Experimental Results and Discussion}

In this section, we clarify the effectiveness of the proposed method through experiments from the aspects of attack resistance and model performance.

\subsection{Setup}
Table \ref{tab_1} lists the fundamental conditions of our experiments.
We virtually configured a server and five clients on a single machine.
For a pre-trained model, we used vit\_small\_patch16\_224, where the patch size $P$ and learning rate $\eta$ were 16 and 0.0001, respectively.
We further set the occurrence probability of zeros $R$ to 0.2, 0.5, and 0.8.
The experiments were conducted using the CIFAR10 dataset, which consists of 50,000 training images and 10,000 test images.
The images were resized from $32 \times 32 \times 3$ to $224 \times 224 \times 3$ by bilinear transformation to correspond to the input size of ViT.

We used 16 images from the CIFAR10 dataset to evaluate robustness against APRIL.
The code of APRIL was referenced from \cite{c16}.
The images were input to the model fine-tuned with CIFAR10, and the inferred images were generated by APRIL from the updated information.

Additionally, we evaluated model performance.
Using a model pre-trained using Image-Net and another model without pre-training, we verified the classification accuracy using the 10,000 test images at the end of each epoch.

\begin{table}[t]
    \centering
    \caption{Fundamental conditions}
    \label{tab_1}
    \begin{tabular}{|c|c|}
        \hline
        \# of clients & 5 \\ \hline
        \hspace{1mm} Pre-trained model \hspace{1mm} & \hspace{1mm} vit\_small\_patch16\_224 \hspace{1mm}  \\ \hline
        \hspace{1mm} Patch size \hspace{1mm} & \hspace{1mm} 16 \hspace{1mm} \\ \hline
        \hspace{1mm} Learning rate \hspace{1mm} & \hspace{1mm} 0.0001 \hspace{1mm} \\ \hline
        \hspace{1mm} Dataset \hspace{1mm} & \hspace{1mm} CIFAR10 \hspace{1mm} \\ \hline
        \hspace{1mm} Occurrence probability of zeros \hspace{1mm} & \hspace{1mm} 0.2, 0.5, and 0.8 \hspace{1mm} \\ \hline
    \end{tabular}
\end{table}

\subsection{Robustness against APRIL}

\begin{figure}[t]
\begin{center}
\begin{minipage}[t]{0.3\linewidth}
\begin{center}
\includegraphics[width=2.2cm]{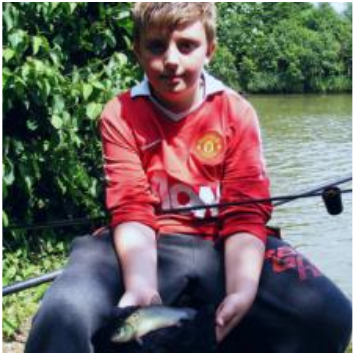}
\end{center}
\subcaption{Training image}
\label{fig:4_1}
\end{minipage}
\begin{minipage}[t]{0.3\linewidth}
\begin{center}
\includegraphics[width=2.2cm]{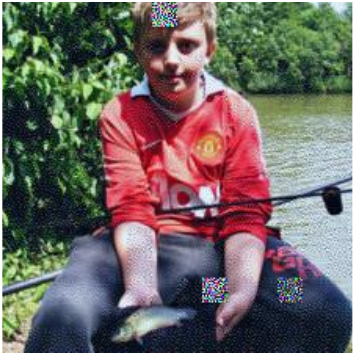}
\end{center}
\subcaption[width=2cm]{Normal learning}
\label{fig:4_2}
\end{minipage}
\begin{minipage}[t]{0.3\linewidth}
\begin{center}
\includegraphics[width=2.2cm]{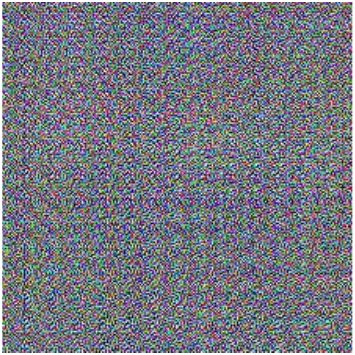}
\end{center}
\subcaption[width=2cm]{Fixed-position $~~~~~$ $~~~~~$ method}
\label{fig:4_3}
\end{minipage}
\end{center}
\begin{center}
\begin{minipage}[b]{0.3\linewidth}
\begin{center}
\includegraphics[width=2.2cm]{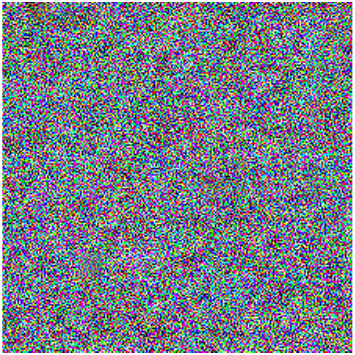}
\end{center}
\subcaption[width=2.2cm]{Proposed method $~$ $~~~~~$ (R=0.2)}
\label{fig:4_4}
\end{minipage}
\begin{minipage}[b]{0.3\linewidth}
\begin{center}
\includegraphics[width=2.2cm]{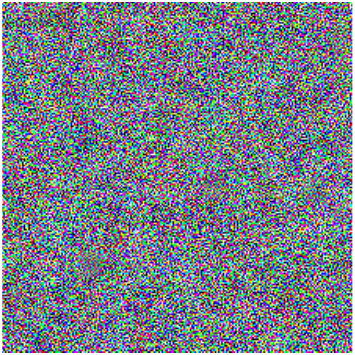}
\end{center}
\subcaption[width=2cm]{Proposed method $~$ $~~~~~$ (R=0.5)}
\label{fig:4_5}
\end{minipage}
\begin{minipage}[b]{0.3\linewidth}
\begin{center}
\includegraphics[width=2.2cm]{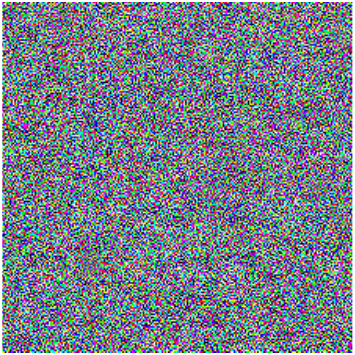}
\end{center}
\subcaption[width=2cm]{Proposed method $~$ $~~~~~$ (R=0.8)}
\label{fig:4_6}
\end{minipage}
\end{center}
\caption{Images inferred by APRIL}
\label{fig:4_1_6}
\end{figure}

Fig. \ref{fig:4_1_6} depicts the restoration results by APRIL using the updated information.
Fig. \ref{fig:4_1_6}\subref{fig:4_1} is a training image, while \ref{fig:4_1_6}\subref{fig:4_2} is an inferred image when using a plain model.
In this case, the visual information of the training image was fully disclosed by APRIL.
In contrast, as shown in Fig. \ref{fig:4_1_6}\subref{fig:4_3}, the fixed-position method prevented APRIL from successful restoration.
Similarly, from Figs. \ref{fig:4_1_6}\subref{fig:4_4}, \subref{fig:4_5}, and \subref{fig:4_6}, which are the images inferred by the proposed method, it was difficult for APRIL to successfully restore the training image even at $R=0.2$.
Thus, we can claim that the proposed method is as robust to APRIL as the fixed-position method.

\subsection{Image Classification Performance}

\begin{figure}[t]
    \centering
    \begin{minipage}[t]{0.9\linewidth}
        \centering
        \includegraphics[width=8cm]{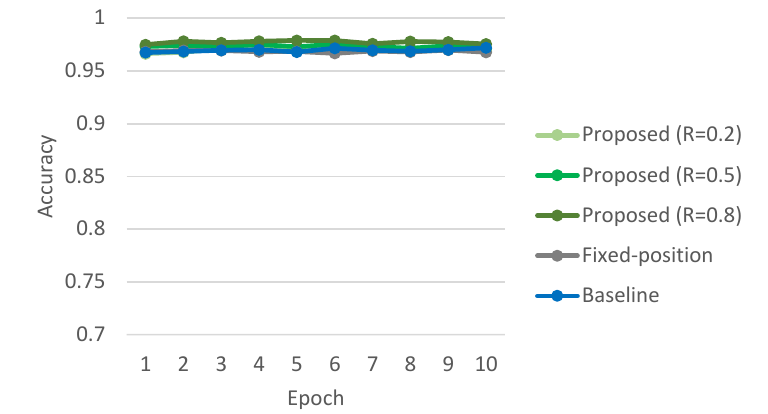}
        \subcaption{W/ pre-training}
    \label{4_g1}
    \end{minipage}
\\
    \centering
    \begin{minipage}[t]{0.9\linewidth}
        \centering
        \includegraphics[width=8cm]{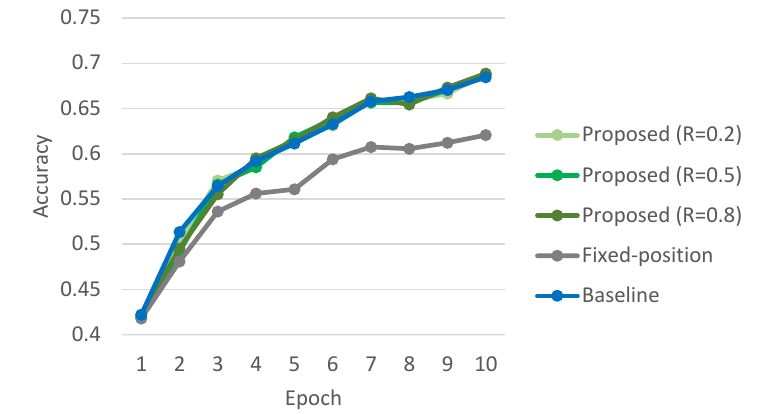}
        \subcaption{W/o pre-training}
    \label{4_g2}
    \end{minipage}
\caption{Comparison of classification accuracy}
\label{4_g}
\end{figure}

We assessed the influence of security enhancement with the proposed and fixed-position methods on classification accuracy.
Fig. \ref{4_g}\subref{4_g1} shows the transition in classification accuracy for each epoch in the case of using the model pre-trained with ImageNet.
In this figure, the baseline is federated learning with the plain model.
It is clear that both the proposed and fixed-position methods retain similar accuracy to the baseline.

Fig. \ref{4_g}\subref{4_g2} is the transition in classification accuracy for each epoch using the model without pre-training.
Although the proposed method could maintain accuracy, the fixed-position method significantly reduced it.
In the case of using a model without pre-training, we should update all layers of ViT because the initial parameters are generally not suitable for classifying desired images.
The fixed-position method does not update the positional embedding layer; this is the reason for the reduction in accuracy.
In contrast, the proposed method updates all the layers, so accuracy can be improved.

\subsection{Discussion}
\begin{figure}[t]
    \centering
    \begin{minipage}[t]{0.9\linewidth}
        \centering
        \includegraphics[width=8cm]{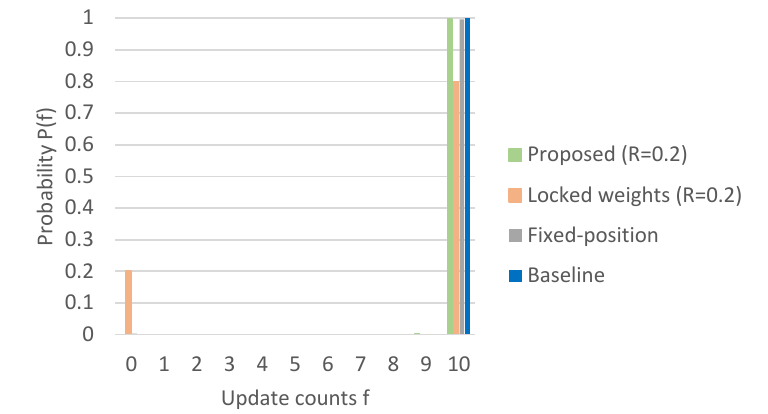}
        \subcaption{$R=0.2$}
    \label{4_g1}
    \end{minipage}
\\
    \centering
    \begin{minipage}[t]{0.9\linewidth}
        \centering
        \includegraphics[width=8cm]{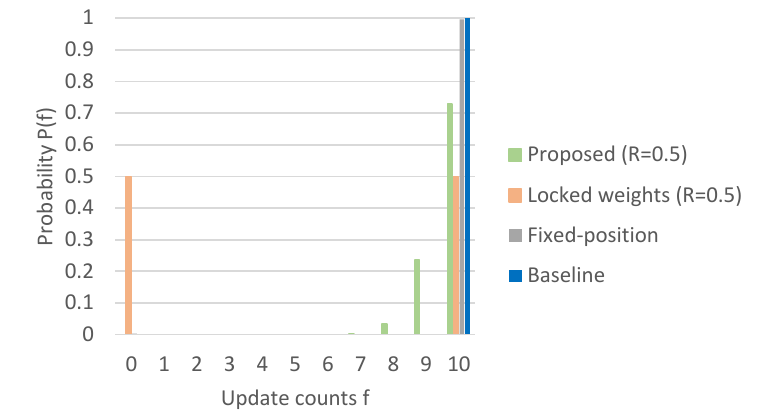}
        \subcaption{$R=0.5$}
    \label{4_g1}
    \end{minipage}
\\
    \centering
    \begin{minipage}[t]{0.9\linewidth}
        \centering
        \includegraphics[width=8cm]{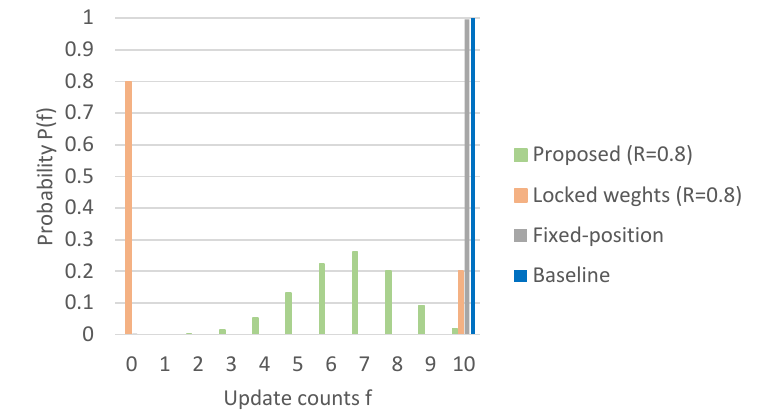}
        \subcaption{$R = 0.8$}
    \label{4_g2}
    \end{minipage}
\caption{Probability distribution of update counts}
\label{fig:g3}
\end{figure}
We discuss why the classification accuracy was not degraded with the proposed method.
If the random binary weight of each client for a given parameter is zero, the parameter is not updated.
Thus, in the case that each client uses a common sequence of random binary weights throughout all epochs, some of the parameters may never be updated.
We call this a locked binary weights method.
In contrast, in the proposed method, the clients use a different sequence of random binary weights varying from epoch to epoch.
This significantly reduces the probability that parameters will never be updated.
In the proposed method, the probability $P(f)$ that a parameter is updated $f$ times is obtained by
\begin{equation}
P(f) = {}_m C_f \times (1-R^{n})^{f} \times R^{n \times (1-f)},
\label{eq:Pf}
\end{equation}
where $m$, $R$, and $n$ denote the number of epochs, the occurrence probability of zeros, and the number of clients, respectively. 
From this equation, the update frequency of each parameter will get larger as the number of clients and epochs increases.
Consequently, the proposed method is expected to train the model more effectively compared with the locked binary weights method.

Fig. \ref{fig:g3} shows the probability distribution of update counts with $m = 10$ and $n = 5$.
Here, in the proposed method and the locked binary weights method, $R$ was defined as $0.2$, $0.5$, and $0.8$.
From the results, it is clear that most of the parameters were constantly updated through $10$ epochs with the fixed-position method.
This method, however, never updates the positional embedding layer.
Updating the positional embedding layer is important for improving the classification accuracy, so the method degrades the accuracy. 
Note that it is difficult to recognize this in Fig.6 since the number of parameters in the positional embedding layer is significantly small.
The proposed method has a large number of updated parameters, although not as large as the fixed-position method.
Since non-updated parameters are not concentrated in a particular layer, our method can also obtain the updates of the positional embedding layer, which is a serious issue with the fixed-position method.
Furthermore, the use of an independent sequence of binary weights for each epoch enabled the proposed method to update more parameters than the locked binary weights method at the values of R defined in this section, that is $R=0.2, 0.5, and, 0.8$.
For these reasons, the proposed method provides not only enhanced security against attacks but also updates of a large number of parameters.
\section{Conclusion}
We proposed a new security enhancement method for ViT-based federated learning. 
With most previous methods, the classification accuracy degrades.
The proposed method not only solves this issue but is also made robust to the APRIL attack by multiplying updated information by random binary weights. 
Through our simulations, we confirmed resistance to APRIL and retention of model performance.
Our future work involves the application of this method to other models such as CNN models.

\end{document}